\documentclass 
[journal]{IEEEtran}

\IEEEoverridecommandlockouts
\usepackage{ifpdf}
\usepackage{cite}
\ifCLASSINFOpdf
   \usepackage[pdftex]{graphicx}
 \DeclareGraphicsExtensions{.eps}
\else
   \usepackage[dvips]{graphicx}
  \DeclareGraphicsExtensions{.eps}
\fi
\usepackage[cmex10]{amsmath}
\usepackage{algorithmic}
\usepackage{algorithm} 
\usepackage{algorithmic}
\usepackage{array}
\usepackage{mdwmath}
\usepackage{mdwtab}
\usepackage{eqparbox}
\usepackage{algorithm}
\usepackage{algorithmic}

\usepackage{mdwmath}
\usepackage{mdwtab}
\usepackage{amsmath}

\usepackage{amssymb}
\usepackage{dsfont}
\PassOptionsToPackage{bookmarks={false}}{hyperref}

\usepackage{fancyhdr}
\begin{document}
\title{Adaptive Modulation in OSA-based Cognitive Radio Networks}
\author{\IEEEauthorblockN{F.~Foukalas$^{*}$, \textit{Member, IEEE,} G.~Karetsos$^\ddag$, \textit{Member, IEEE,} G.~
G.K.~Karagiannidis$^\dag$} \textit{Senior Member, IEEE,}\\
\IEEEauthorblockA{$^*$Informatics and Telecommunications, University of Athens, Athens, Greece\\
$^\ddag$Information and Communication Technologies, TEI of Larissa, Greece\\
$^\dag$Electrical and Computer Engineering, Aristotle University of Thessaloniki, Thessaloniki, Greece\\}}

\maketitle

\begin{abstract}
Opportunistic spectrum access is based on channel state information and can lead to important performance improvements for the underlying communication systems. On the other hand adaptive modulation is also based on channel state information and can achieve increased transmission rates in fading channels. In this work we propose the combination of adaptive modulation with opportunistic spectrum access and we study the anticipated effects on the performance of wireless communication systems in terms of achieved spectral efficiency and power consumption.       
\end{abstract}

\begin{keywords}
adaptive modulation, opportunistic spectrum access, cognitive radio networks, spectral efficiency gain.   
\end{keywords}

\IEEEpeerreviewmaketitle
\section{Introduction}
Adaptive modulation is nowadays one of the key components in most wireless communication standards such as the high speed packet access (HSPA) and the worldwide interoperability for microwave access (WiMAX), since it allows for high data rates over fading channels ~\cite{c1}. In particular the transmission rate is adapted based on the channel conditions
that are estimated at the receiver’s side and are made available also at the transmitter throughout a feedback channel. When adaptive modulation is implemented in conjunction with power control at the physical layer then a variable rare variable power (VRVP) modulation is considered ~\cite{c2}. Two alternative schemes of VRVP have been specified, known as continuous rate (CR) and discrete rate (DR) although the latter is more practical from implementation point of view.

\let\thefootnote\relax\footnote{This paper accepted to to IEEE Vehicular Technology Conference 2011 (Wireless Networks, 49963)}

On the other hand, cognitive radio (CR) has been recently proposed for enhancing spectrum utilization of licensed wireless networks, also known as primary networks (PN), when certain conditions apply ~\cite{c3}. The underutilized or unused spectrum resources can be exploited by the so called cognitive or secondary networks (SN) as long as their operation
is not harmful for the PN. Two main types of cognitive radio networks (CRNs) have been identified so far, namely opportunistic spectrum access (OSA) and spectrum sharing (SS) ~\cite{c4}. The first one relies on exploiting spectrum gaps being available in the PN which are recognized by the SN via spectrum sensing and the second one relies on the coordinated sharing of a spectrum band among the PN and the SN. In addition if an SS CRN employs also spectrum sensing, then the specific CRN is known as sensing-based spectrum sharing CRN and can be assumed as a third type of CRNs ~\cite{c8}. All three CRN types exploit channel state information (CSI) in order to provide enhanced spectral efficiency over the considered wireless channels ~\cite{c5} ~\cite{c7}. To this end, one of the main techniques employed is power control which regulates the transmission of the SN users while protecting the PN ones ~\cite{c6}.    

In opportunistic spectrum access (OSA) systems cognitive users (CUs) are able to detect and use channels that have been originally allocated to primary users (PUs) when these are recognized as being idle ~\cite{c4}. The knowledge of the channel’s state is very important for allocating idle channels. In particular it allows an optimal power allocation on a specific channel ~\cite{c9}. Furthermore, it can lead to the selection of the most appropriate modulation scheme when adaptive modulation is being considered ~\cite{c1}. Both OSA and adaptive modulation are heavily depending on the channel state and thus their deployment could be jointly considered. In this letter, we study the incorporation of adaptive modulation in OSA systems. Specifically, by assuming a spectrum pooling, as the OSA system ~\cite{c7} ~\cite{c10}, we derive the gain achieved in spectral efficiency and obtain the optimal power allocation from the application of different kinds of adaptive modulation. The obtained numerical results denote the achievable performance gain in spectral efficiency and the power requirements from such a combination.

The rest of this paper is organized as follows. Section 2 describes the corresponding system model when adaptive modulation is implemented in OSA-based CRNs. Section 3 provides the performance analysis of adaptive modulation in OSA-based CRNs over fading channels. In section 4, we present and discuss the obtained numerical results and in section 5 we provide the conclusions.

\section{Opportunistic Spectrum Access CRNs System Model} \label{system}

\begin{figure*}
\begin{center}
  \includegraphics[width=6in]{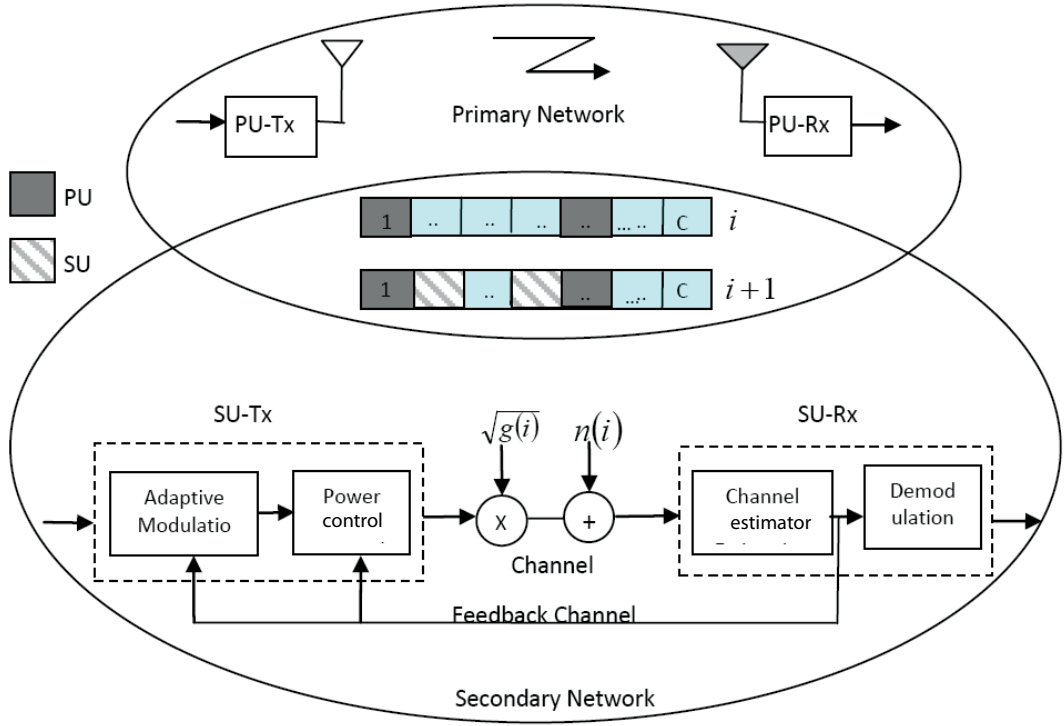}
\end{center}
  \caption{System model of Opportunistic Spectrum Access CRN}
  \label{fig:systemmodel}
\end{figure*}

We assume an OSA CRN with $c \in C$ channels and $u \in U$ users where each user $u$ is served relying on a spectrum pooling strategy that first serves the PU and subsequently the SUs ~\cite{c7}. Fading channels are assumed with channel gain $g(i)$ and additive white Gaussian noise (AWGN) $n(i)$, both at time $i$. The average transmit power over the fading channel is $P$, the AWGN is with power density $N_0/2$ and the received bandwidth is $B$. An SU can access a channel $c$ if and only if a predefined level on the instantaneous transmit power $P$, is achieved. This level is determined from the channel state information (CSI) which represents the received Signal-to-Noise-Ratio (SNR), $\gamma$ that is equal to $g(i)P/N_0B$ at time $i$ and for a unit of bandwidth. Thus, the transmit power $P$ is controlled by $\gamma$, and we denote it as $P(\gamma)$. This policy is known as optimal power allocation in wireless communications and relies on the channel state estimation $g^^(i)$ at the receiver side
~\cite{c11}. The channel estimation i.e. SNR $\gamma$ is also available at the transmitter side via a feedback channel. We assume that the CSI is perfectly available at the receivers i.e. PU-Rx, SU-Rx and that the feedback channel does not induce any delays on the CSI’s transmission. Besides, a set of $M-ary$ Quadrature Amplitude Modulations (M-QAM) is considered, that their selection is governed from the CSI. Thus, the combined system first determines if a user u can access a channel c and in the sequel it selects the transmission rate $R = log2(M)$ via the selection of the appropriate $M-ary$ modulation from the signal
set $M$ according to the estimated CSI. Fig.1 shows the system model of the considered OSA CRN. The model shows one PU and one SU and the channels C which are available for access from both user types at time $i$ and $i + 1$ respectively. For instance, at time the PU occupies two channels in which the specified level in SNR has been achieved and in the same way the SU occupies two different channels in the considered spectrum at time $i + 1$.


Furthermore, we make the following assumptions for the considered system: a) the transmission of each symbol is accomplished with a symbol period $Ts = 1/B$ using ideal raised cosine pulses and b) the fading channel is varying slowly in time, i.e. the receiver is able to sense and track the channel fluctuations and thus it corresponds to a block flat fading channel model with an average received SNR, $\bar{\gamma}$ ~\cite{c12}.

\section{Adaptive Modulation in Opportunistic Spectrum Access CRNs over Fading Channels} \label{system}

\subsection{Continuous Rate adaptive modulation in OSA CRN} \label{system1}
The continuous rate (CR) adaptive modulation with MQAM constellation set results in the following expression for the constellation set for a specific bit error rate (BER) ~\cite{c1}

\begin{eqnarray} \label{eq1}
 M(\gamma) = 1 + \frac{1.5\gamma}{-ln(5BER)} \frac{5\gamma}{\bar{P}}P(\gamma)
\end{eqnarray}

where $\gamma$ is the received SNR. On the other hand, the aim in OSA CRNs is to allocate a channel $c$ to user $u$ which maximizes the average spectral efficiency (ASE) \footnote[1]{The term average spectral efficiency is used when fading channels are
assumed} given as follows ~\cite{c7}

\begin{eqnarray} \label{eq2}
\nonumber S_e &=& E[log_2 M(\gamma)] \\
&=& \int log_2(1+\frac{1.5\gamma}{-ln(5BER)}\frac{P(\gamma)}{P}) d\gamma 
\end{eqnarray}

subject to the following power constraint

\begin{eqnarray} \label{eq3}
 \int P(\gamma)p(\gamma)d\gamma = \bar{P} 
\end{eqnarray}

Thus, the optimal power allocation which maximizes the ASE in OSA CRNs with CR adaptive modulation is given as follows ~\cite{c2}

\begin{eqnarray}\label{eq4}
\frac{P(\gamma)}{\bar{P}}=
\begin{cases}
\ \frac{1}{\gamma_0}-\frac{1}{\gamma K}, \gamma \geq \frac{\gamma_0}{K} \\
\ 0 , \gamma < \frac{\gamma_0}{K}\\ 
\end{cases}
\end{eqnarray}

where $K$ is an effective power loss that retains the BER value and it is equal to

\begin{eqnarray} \label{eq5}
 K = \frac{-1.5}{ln(5BER)} 
\end{eqnarray}

Substituting (\ref{eq4}) into (\ref{eq2}), the ASE for the CR MQAM is maximized up to a cut-off level in SNR denoted as $\gamma_k=\gamma_0/K$ and thus it is obtained as follows

\begin{eqnarray} \label{eq6}
 \langle S_e \rangle_{CR} = \int_{\gamma_k}^{\infty} log_2(\frac{\gamma}{\gamma_K})p(\gamma)d\gamma 
\end{eqnarray}

In the considered OSA CRN, equation (\ref{eq4}) gives the ASE that the PU achieves at a channel $c$ denoted as $Se_{1,c}$ since it is served first from the OSA strategy ~\cite{c7}. Thus, the achieved ASE by a SU $u$ denoted as $Se_{u,c}$ is equal to

\begin{eqnarray} \label{eq7}
 Se_{u,c} = \Delta_{1,c}Se_{1,c} 
\end{eqnarray}

where $\Delta_{1,c}$ is the spectrum factor gain which represents the probability that the channel c is not occupied by the PU. This gain depends on the cut-off level in SNR  $k$ of the optimal power allocation over the fading channel and hence it is obtained as follows

\begin{eqnarray} \label{eq8}
 \Delta_{1,c} = \int_{0}^{\gamma_k} p(\gamma)d\gamma 
\end{eqnarray}

If we generalize this strategy for $U$ users, the sum ASE which an OSA CRN provide is given as follows

\begin{eqnarray} \label{eq9}
 Se_{sum} = \Sigma_{u=1}^{U}Se_{u,c} = = \Sigma_{u=1}^{U}\Delta_{1,c}Se_{1,c} =  \frac{1-\Delta_{1,c}^{U}}{1-\Delta_{1,c}}Se_{1,c} 
\end{eqnarray}

The term $1-\Delta_{1,c}^{U}/1-\Delta_{1,c}$ is called total band factor gain and it represents the percentage of the channels that are remained unused and thus they can be used by the SUs which are served with a specific priority by the OSA CRN.

\subsection{Discrete Rate adaptive modulation in OSA CRN} \label{system2}
We now consider a discrete rate (DR) MQAM with a constellation set of size $N$ with $M_0 = 0$,$M_1 = 2$ and $M_j = 2^{2(j-1)}$ for $j = 2,...,N$. At each symbol time, the system transmits with a constellation from the set ${M_j = 0,1,...,N}$ ~\cite{c2}. The choice of a constellation depends on the $\gamma$ fade level i.e. SNR over that symbol time while the $M_0$ constellation corresponds to no data transmission. Therefore, in OSA CRNs, for each value of  $\gamma$, the SU-Tx decides which constellation $M$ to transmit and what is the associated transmit power $P$ in order to maximize the average spectral efficiency (ASE). The ASE is now defined as the sum of the data rates of each constellation multiplied with the probability that this constellation will be selected and thus it is given as follows

\begin{eqnarray} \label{eq10}
 \langle Se \rangle_{DR} = \Sigma_{j=1}^{N}log_2(M_j)p(\gamma_i\leq\gamma\leq\gamma_{i+1})
\end{eqnarray}

subject to the following power constraint

\begin{eqnarray} \label{eq11}
 \Sigma_{j=1}^{N} \int_{\gamma_j}^{\gamma_{j+1}} \frac{P_j(\gamma)}{\bar{P}} p(\gamma)d\gamma = 1
\end{eqnarray}

where $P_j(\gamma)/\bar{P}$ is the optimal power allocation that is obtained from (\ref{eq3}) for each constellation $M_j$ with a fixed BER as follows

\begin{eqnarray}\label{eq12}
\frac{P_j(\gamma)}{\bar{P}}=
\begin{cases}
\ (M_j-1)\frac{1}{\gamma_K} -\frac{1}{\gamma K}, M_j \leq \frac{\gamma}{\gamma^*} \leq M_{j+1}\\
\ 0 , M_j=0\\
\end{cases}
\end{eqnarray}

where $\gamma^*$ is the cut-off level in SNR of the optimal power allocation which optimize the amount of the fading regions $\gamma_j$ for $j = 0,1,...,N$ according to $\gamma_j=\gamma^*M_j$ and thus the maximization of the spectral efficiency is being accomplished. Therefore, the band factor gain and the sum ASE in OSA CRNs which implement a DR adaptive modulation depends on this the cut-off level in SNR, $\gamma^*$  and they are obtained by equations ~\cite{c8} and ~\cite{c9} accordingly.

\section{Numerical Results}
In Fig.2 we present the results obtained when continuous rate (CR) adaptive modulation in OSA CRN over fading channel is considered. We assume a Rayleigh distribution for the fading channel with probability density function equal to $1/\bar{\gamma}exp(-\gamma/\bar{\gamma})$ where $\gamma$ and $\bar{\gamma}$ are the instantaneous and the average received SNR respectively. We depict the results for bit error rates (BER) equal to $10^{-3}$ and $10^{-6}$ respectively. With solid lines are shown the results when only the PU is considered, which is the case of a conventional network i.e. which does not serve any SUs. With dashed lines are shown the results obtained for the OSA CRN with a number of users equal to $U = 5$. An important performance gain in the OSA CRN is observed in comparison with the performance of the conventional network. In detail, for both BER cases the additional ASE is close to $0.5 bits/sec/Hz$ at low average SNR regions e.g. $ \bar{\gamma}= 0dB$. Besides, the additional ASE is close to $0.3 bits/sec/Hz$ at moderate average SNR regions e.g. $\bar{\gamma}= 10dB$ and finally the ASE is close to $0.1bits/sec/Hz$ at high average SNR regions e.g.  $ \bar{\gamma}= 20dB$. This behavior in particular is explained from the fact that the probability that a channel is not allocated to the PU is larger for the low average SNRs. In other words, for the low average SNR regions, the cut-off level in SNR  $K$ is getting higher and in consequence the band factor gain in equation (\ref{eq8}) is getting higher too. The opposite is applied for the high average SNRs where the PU is more likely to transmit on the channel or a unit bandwidth in general since the cut-off level in SNR  $K$ is getting lower and thus the constraint for allocating a channel is relaxed. Fig.3a and Fig.3b show the results obtained when the discrete rate (DR) adaptive modulation in OSA CRN over a Rayleigh fading channel is considered. We assume the aforementioned Rayleigh distribution with  to be the average received SNR. We depict the results for a bit error rate (BER) equal to and $10^{03}$ in Fig.3a and the results for a BER equal to 10􀀀6 in Fig.3b. With solid lines are shown the results for the conventional network i.e. $U = 1$ and with dashed lines are shown the results obtained for the OSA CRN with a number of users equal to $U = 5$. We consider different fading regions i.e. 5 fading regions with a set of MQAM constellations {0,2,4,16, 64} i.e. 4 fading regions with a set of MQAM constellations {0,2,4,16} and 3 fading regions  with a set of MQAM constellations {0,2,4}. Again, the performance gain is remarkable for low average SNR regions for the same reasons as with the CR OSA CRN. It should be noticed that the performance gain is identical at low average SNR regions for all fading regions i.e. 5,4 and 3 that is close to $0.3bits/sec/Hz$. On the other hand, the performance gain is negligible at high average SNR regions. Regarding the different BER values, the tighter the BER is get i.e. $BER = 10^{-6}$ , the larger the performance gain is become, something that we discuss in detail in Fig.5 which illustrates the total band factor gain for the CR and DR implementations of adaptive modulation in OSA CRNs.

\begin{figure}
  \includegraphics[width=95mm,height=70mm]{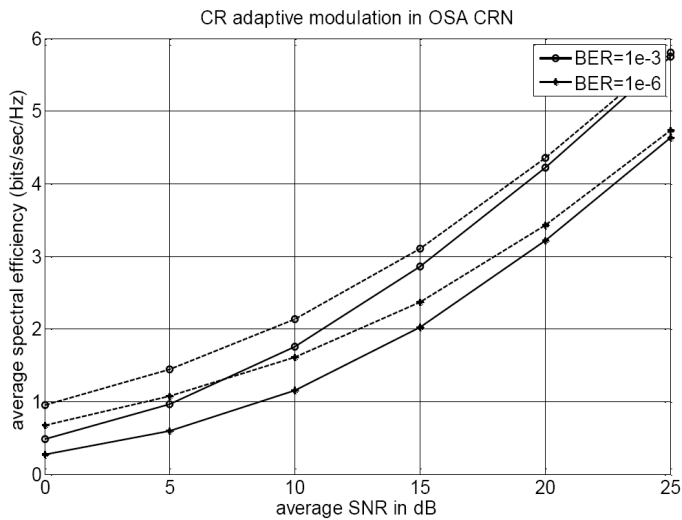}\\
  \caption{Average spectral efficiency of CR adaptive modulation in OSA CRN over Rayleigh fading channel}
  \label{fig:2}
\end{figure}

\begin{figure}
  \includegraphics[width=\columnwidth]{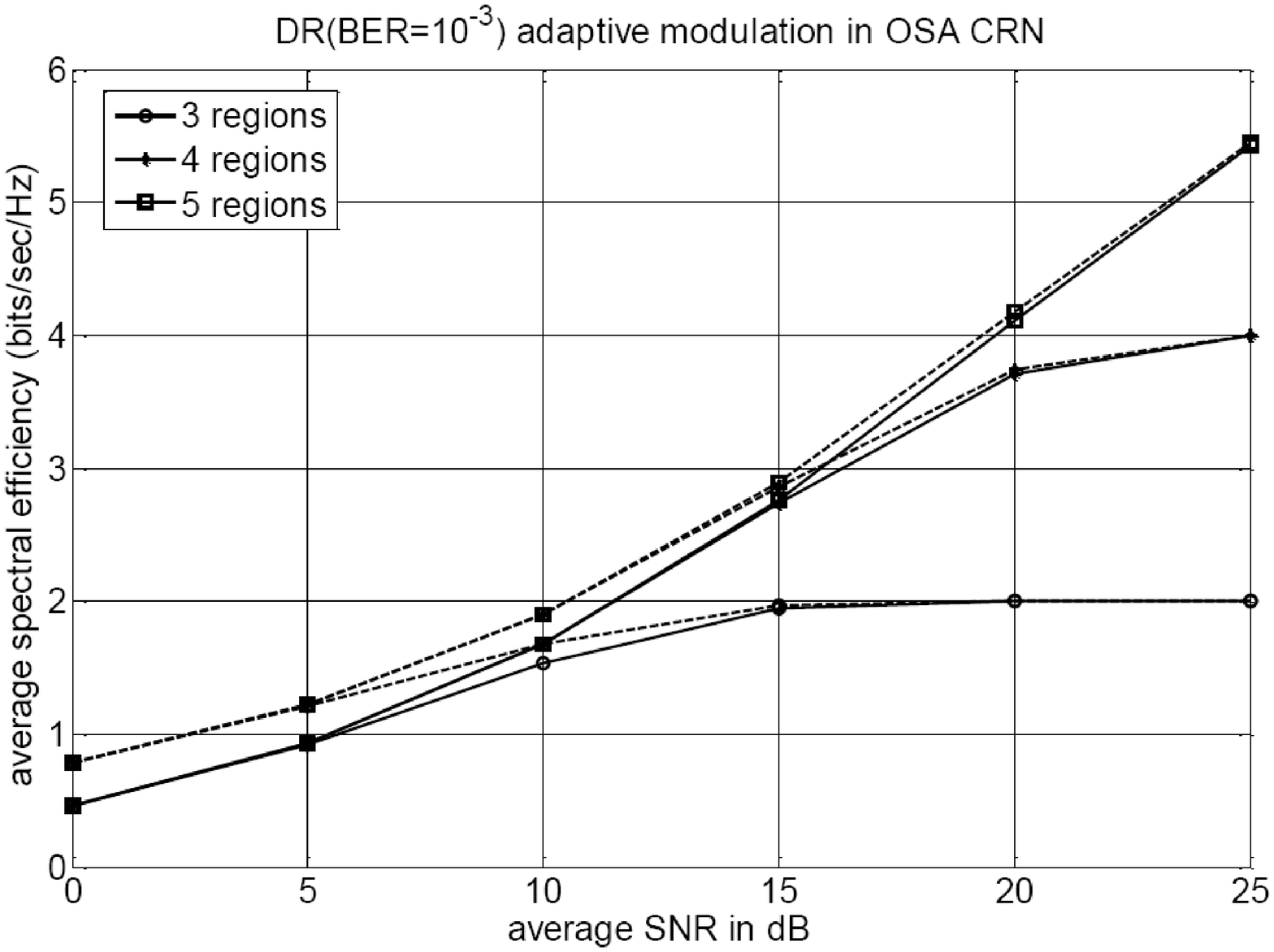}\\
  \caption{Average spectral efficiency of DR ($BER = 10^{-3}$) adaptive modulation in OSA CRN over Rayleigh fading channel}
  \label{fig:3a}
\end{figure}

\begin{figure}
   \includegraphics[width=\columnwidth]{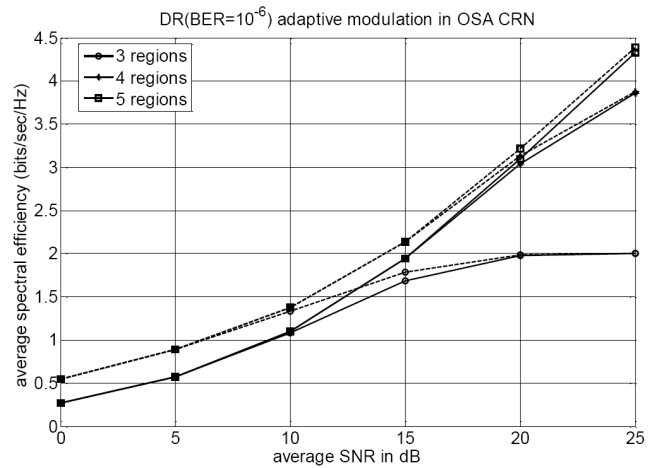}\\
  \caption{Average spectral efficiency of DR ($BER = 10^{-6}$) adaptive modulation in OSA CRN over Rayleigh fading channel}
  \label{fig:3b}
\end{figure}

Fig.4 shows the total band factor gain of adaptive modulation in OSA CRN as it is obtained from equation (\ref{eq9}). We depict both the cases of CR and DR adaptive modulation over a Rayleigh fading channel. With solid lines are shown the results obtained for a BER equal to $10^{-3}$ and with dashed lines are shown the results obtained for a BER equal to $10^{-6}$. The OSA CRN is considered with a number of users equal to $U = 5$. Notably, the largest total band factor gain is achieved in case of CR adaptive modulation with a BER equal to $10^{-6}$ and the smallest one is achieved in case of DR adaptive modulation with 3 regions and a BER equal to 10􀀀3 . Therefore, the tighter the BER criterion is become, the larger the advantage of the application of the OSA strategy in CRNs. We should further notice that the gain of the CR and DR adaptive modulations with a high number of regions i.e. 5 was expected due to the transmission with high bit rates in terms of bits per symbol and in consequence with a high average spectral efficiency.

\begin{figure}
   \includegraphics[width=\columnwidth]{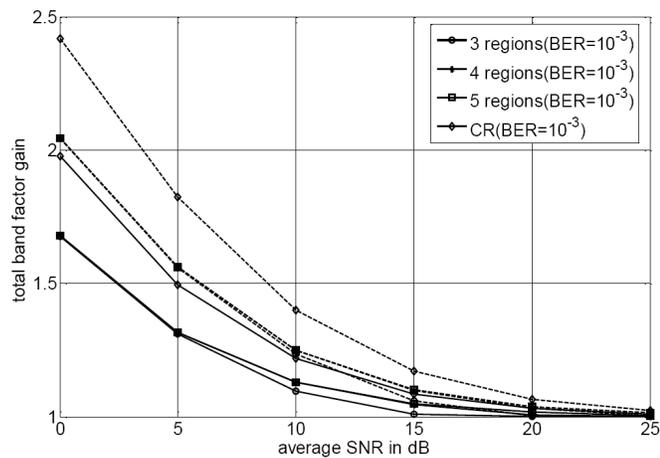}\\
  \caption{Total band factor gain of adaptive modulation in OSA CRN}
  \label{fig:4}
\end{figure}

\begin{figure}
  \includegraphics[width=\columnwidth]{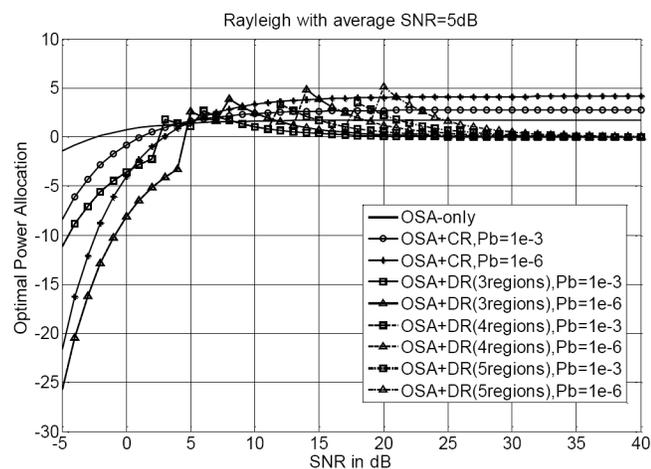}\\
  \caption{Optimal power allocation for different adaptive modulation schemes
versus the received SNR for a $5db$ average transmit power}
  \label{fig:5}
\end{figure}

The optimal power allocation versus the received SNR for an average transmit power of 5db is shown in Fig.5. This value is selected from Fig. 4 since we observe that an important gain in spectral efficiency is achieved in this average SNR region. By decreasing the number of modulations in the adaptive scheme as well as the upper bound of the error probability, the consumed power is decreasing. On the other hand, the most power demanding case is when no modulations are being used.

\section{Conclusion}
In this work, the incorporation of adaptive modulation in
opportunistic spectrum access cognitive radio networks over
fading channels is studied. In particular we have shown that
the usage of adaptive modulation in OSA systems leads
to an increased spectral efficiency and to decreased power
consumption. This improvement is getting better when lower
number of regions and tighter error probabilities are applied.
Future work includes the assessment of spectral efficiency
when adaptive modulation is considered for spectrum sharing
cognitive radio networks with or without sensing information
availability.

\end{document}